\documentclass{emulateapj}

\newcommand{\D}[1]{\, d #1 \,}
\newcommand{\vc}[1]{{\bf#1}}
\newcommand{\vch}[1]{\hat{\bf#1}}
\newcommand{\avg}[1]{\left< #1 \right>}
\newcommand{\ft}[1]{\widetilde{#1}}

\newcommand{\eps}{\varepsilon}

\usepackage{verbatim}
\usepackage{amsmath, amsthm, amssymb}
\def\kms{km~s$^{-1}$}
\def\ga{\mathrel{\hbox{\rlap{\hbox{\lower4pt\hbox{$\sim$}}}\hbox{$>$}}}}
\def\la{\mathrel{\hbox{\rlap{\hbox{\lower4pt\hbox{$\sim$}}}\hbox{$<$}}}}

\shorttitle{Velocity Coordinate Spectrum of the SMC}

\shortauthors{CHEPURNOV ET AL. }
\begin{document}

\title{The Turbulence Velocity Power Spectrum  of Neutral Hydrogen in the Small Magellanic Cloud}

\author{ A. Chepurnov\altaffilmark{1},  B. Burkhart\altaffilmark{2}, A. Lazarian\altaffilmark{1}, S. Stanimirovic\altaffilmark{1} }
\altaffiltext{1}{Astronomy Department, University of Wisconsin, Madison, 475 N. 
Charter St., WI 53711, USA}
\altaffiltext{2}{Harvard-Smithsonian Center for Astrophysics, 60 Garden St., Cambridge, MA 0213}

\begin{abstract}

We present the results of the  Velocity Coordinate Spectrum (VCS)  technique to calculate the velocity power spectrum of turbulence in the Small Magellanic Cloud (SMC) in 21cm emission.
We have obtained a velocity spectral index of -3.85 and an injection scale of 2.3 kpc.
The spectral index is steeper than the Kolmogorov index which is expected for shock-dominated turbulence which is in agreement with past works on the
SMC gas dynamics.  The injection scale of 2.3 kpc suggests that tidal interactions with the Large Magellanic Cloud are the dominate driver of turbulence in this dwarf galaxy.  This implies turbulence maybe driven by multiple mechanisms in galaxies in addition to supernova injection and that galaxy-galaxy interactions may play an important role.

\end{abstract}
\keywords{ISM: structure --- MHD --- turbulence,  SMC}

\section{Introduction}
\label{intro}

Magnetohydrodynamic (MHD) turbulence is ubiquitous in the interstellar medium (ISM) of galaxies.  It is an essential element in many physical processes, such as the formation of stars (Federrath \& Klessen 2012; Burkhart, Collins \& Lazarian 2015)
and transport of cosmic rays (Lazarian \& Yan 2014), and is a large fraction of the energy budget of the Milky Way (see Elmegreen \& Scalo 2004). 
The evidence for the existence of turbulence in the ISM is overwhelming and includes 
the ``Big Power Law'' and ``Extended Power Law'' of the electron density fluctuations in  Armstrong et al. (1994) and Chepurnov \& Lazarian (2010), respectively,
fractal structure of molecular clouds (Stutzki et al. 1998), 
and intensity fluctuations and broadened linewidths in channel maps (see Crovisier \& Dickey 1983, Green 1993, Stanimirovic et al. 1999, 
Deshpande, Dwarakanath \& Goss 2000).  Despite the importance of ISM turbulence for a wide range of physical processes in galaxies, little is known about it due to the difficulty of measuring the velocity power spectrum of turbulence which provides information on the injection and dissipation scales, energy amplitude, and spectral slope.

For studies of turbulence, velocity fluctuations reflected in position-postion-velocity (PPV) radio cubes
are critical as they trace the underlying fluid behavior. Along with density fluctuations  they contribute  to the intensity 
fluctuations in PPV space.  As such, many techniques for analyzing PPV fluctuations in the context of turbulence have been proposed including
velocity centroids (Esquivel \& Lazarian 2005a,b; Ossenkopf et al. 2006; Esquivel et al. 2007),
hierarchical tree methods (Goodman et al. 2009; Burkhart et al. 2013),
principal component analysis (Heyer \& Schloerb 1997)
and the $\Delta$ variance technique (Ossenkopf et al. 2008a,b)
However, these techniques face difficulty to reliably separate out the underlying velocity and density contributions in PPV space in the supersonic limit.  This is essential for obtaining the velocity power spectrum of turbulence in astrophysical settings.

For the purpose of obtaining the velocity power spectrum the Velocity Channel Analysis (VCA, Lazarian \& Pogosyan 2000, 2004, henceforth LP00, LP04) and Velocity Coordinate Spectrum (VCS, see Lazarian \& Pogosyan 2000, 2006, 2008,
henceforth, LP00, LP06, LP08) have been developed based on analytic expression for the behavior of the correlation function (which is isomorphic to the power spectrum) in PPV space. A concise review of these
extensive works is given in Lazarian (2009). 

The VCS and VCA techniques are entirely data driven (i.e. comparisons or scaling to numerical simulations are not required) and  are complimentary to each other as the success of the VCA application depends primarily on the spatial resolution while the VCS depends primarily on the spectral resolution.
This is because the  VCA technique is concerned with taking the spatial power spectrum of the 
two-dimensional slices of PPV while the VCS studies the fluctuations of PPV intensities along the velocity coordinate of the PPV cube.
Both techniques have been applied to multi-wavelength observations as well as undergone numerical tests (Stanimirovic \& Lazarian 2001; Chepurnov et al. 2010; Chepurnov et al. 2009)

In this paper we apply the VCS to the Small Magellanic Cloud (SMC) in 21cm emission.
The SMC is a dwarf irregular galaxy in the Local Group which has a gas rich ISM environment and is metal poor (see, Stanimirovic et al. 1999, henceforth known as SX99).  The SMC is nearby (60 kpc; Westerlund 1991) and is close enough to have resolved scales HI from $\approx$ 4.5kpc down to 30pc.  This is in contrast to clouds in the  Milky Way where distance 
determination is relatively uncertain and special geometrical considerations must be taken into account for application of the VCS as was done in the application of the VCS
in high latitude HI
data (Chepurnov et al. 2010).
Another advantage of investigating the SMC is that it has been used as a test bed for numerous turbulence studies.
HI observations of the SMC, obtained using the Australia Telescope Compact Array (ATCA) and 
the Parkes telescope (SX99),
have already been studied with the VCA technique (Stanimirovic \& Lazarian 2001) as well as the traditional power spectrum. Other statistics were applied to the SMC by Burkhart et al. (2010) which provide information on compressibility, (i.e. sonic Mach number) such as the PDFs and spatial power spectrum.
In addition, the topology of the SMC and its relation to the turbulence environment was quantified using the 
genus statistic in Chepurnov et al. (2008).
Because the SX99 SMC data set is so well studied, it is a perfect candidate to 
test and compare the newer VCS technique. 
Thus by studying this data set we can acquire new information, test and confirm past results,
and  validate the promise of these tools for further use in other observational studies. 

This paper is organized as follows: in Section \ref{sec:smc}  we describe the SMC HI data set used, in Section  \ref{sec:vcs} we review the VCS technique, 
 in Section \ref{sec:res} we detail the results of our analysis and finally in Sections and \ref{sec:disc} and \ref{sec:con} 
we discuss the implications of our findings and draw our conclusions.

\section{The SMC in HI: Test Galaxy for Turbulence Studies}
\label{sec:smc}

\begin{figure*}[tbh]
\centering
\includegraphics[scale=1.0]{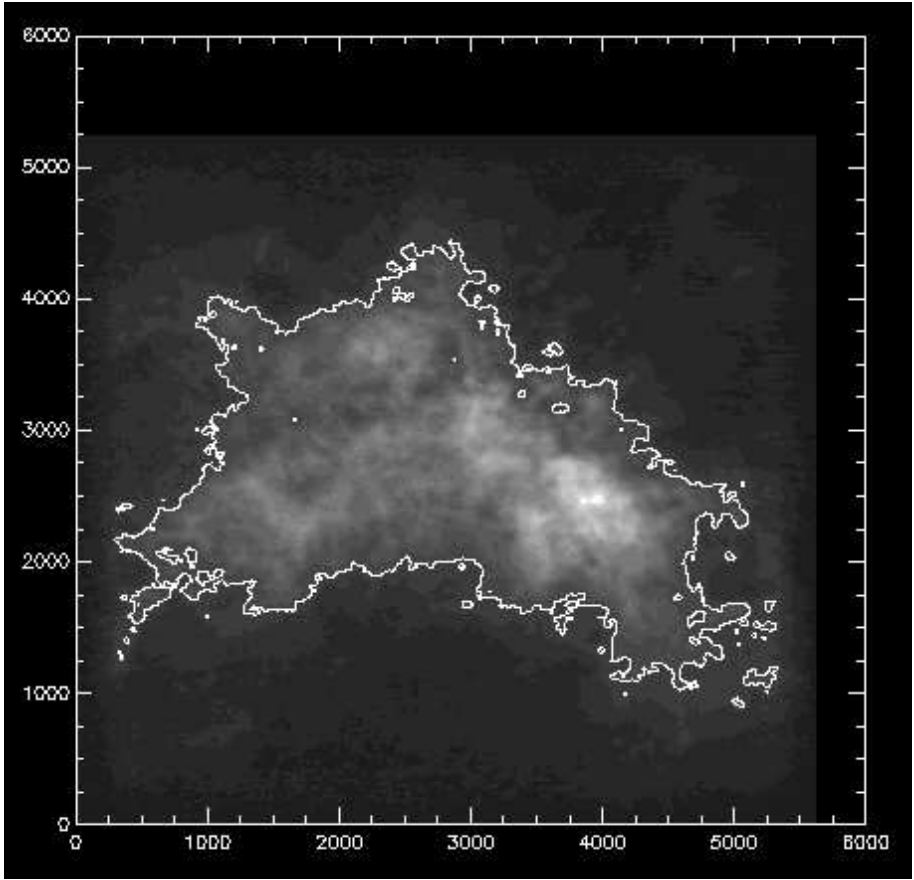}
\caption{SMC integral intensity map, with the contour surrounding region with 
the signal level being more than 20\% from the maximal intensity, selected for the VCS computation. The spatial scaling is in parsecs.}
\label{fig:smc}
\end{figure*}

The SMC HI data set used in our analysis combines both single dish (Parkes, Stanimirovic et al. 1999) and 
interferometer (ATCA, Staveley-Smith et al. 1997)
Both sets of observations were then combined
(see Stanimirovic et al. 1999), resulting in the final HI data
cube with angular resolution of 98$''$, velocity resolution of 1.65 \kms, 
and 1-$\sigma$ brightness temperature sensitivity of 1.3 K.  The data has a continuous range of spatial scales between 30 pc and
4 kpc. The velocity range of the observations
is 90-215 \kms. For more details about the ATCA and Parkes
observations, data processing, and data combination 
(short spacings correction) see Staveley-Smith et al. (1997) and Stanimirovic et al. 1999.
 HI self-absorption we corrected for following the scheme of Stanimirovic et al. 1999.
The correspondent HI intensity image is shown in Figure ~\ref{fig:smc}.

\section{Overview of the VCS Technique}
\label{sec:vcs}

The suggestion to analyze the data calculating the spectrum along the velocity axis of PPV cubes was first made in LP00 and elaborated in LV06. It was only later that the technique was applied to actual observational data (see Padoan et al. 2009, Chepurnov et a. 2010).

LP06 mostly dealt with the asymptotic scaling of turbulence and the observational study by Padoan et al. 2009 is an illustration of the application of the VCS to molecular clouds. The VCS was applied to galactic HI data in Chepurnov et al. (2010)  not only to obtain the spectral slope, but also the injection scale of the turbulence as well as other pieces of useful information. Here we will review the main points of the VCS technique following the way as it is presented in Chepurnov \& Lazarian (2009).

\begin{figure*}[tbh]
\centering
\includegraphics[scale=1.2]{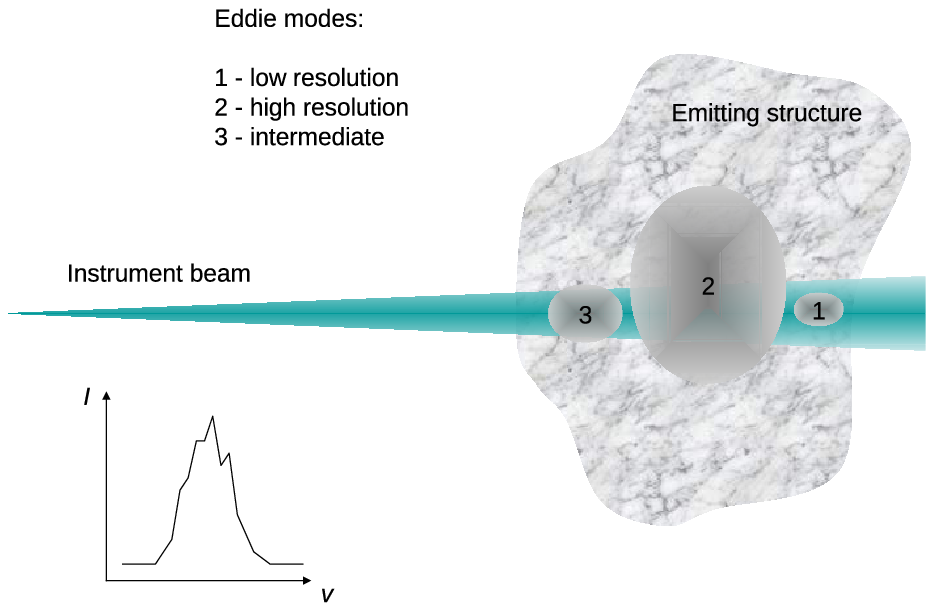}
\caption{The relation between the VCS technique and the effective beam resolution.  Eddies that lie resolved within the telescope beam are 
labeled 1 and are considered part of the ``low-resolution'' mode, eddies with size larger than the telescope beam are labeled 2 and are considered part of the ``high resolution mode''. }
\label{fig:vcs}
\end{figure*}

Below we briefly present the idea  of the VCS technique. A more detailed discussion, including the analysis of the effects of absorption is given in 
Lazarian \& Pogosyan (2006). The contributions to the  PPV power spectrum is
are coming  the power spectrum of velocity, $P_v\propto k^{-\alpha_v}$, and the power spectrum of emissivity, $P_{\varepsilon}\propto k^{-\alpha_{\eps}}$, where $k$ is the wavevector and $k\propto 1/l$ where $l$ is the spatial scale
Power spectra $P_v$ and $P_{\eps}$, respectively, determine the turbulent cascade of energy and the distribution of density in the interstellar medium.
The VCS technique obtains the underlying velocity power spectrum by calculating the 1D Fourier power spectrum of intensity fluctuations (i.e. $P_1(k_v)$) along the velocity axis of a PPV cube. The VCS is based on the theory that relates the fluctuations of intensity in the PPV scale to the underlying fluctuations of velocity and density.
The procedure that we employ is based on the fact that the  $P_1(k_v)$ depends on the angular resolution of the telescope beam (see Figure 1). 
By investigating how $P_1$ changes when changing from high to low angular resolution, and by utilizing the information from intensity map (or column density map) we can evaluate the velocity power spectrum.

The spectral line signal that we deal with is the HI intensity measured at velocity $v_0$ and at a given beam position $\hat{e}$:
\begin{equation} 
\label{eq:sig_rat}
S(\vch{e},v_0)
  = \int w(\vch{e},\vc{r}) \D{\vc{r}} \eps(\vc{r}) f(v_r(\vc{r}) + v_r^{reg}(\vc{r}) - v_0),
\end{equation}
where $\eps$ is normalized emissivity,$f$ denotes the convolution between the spectrometer channel sensitivity function  and the temperature dependent Maxwellian distribution of velocities of emitting particles. In addition, $v_r^{reg}$ is a line-of-sight component of the regular galactic velocity shift, $v_r$ is the line-of-sight component of the random turbulent velocity.  The window function $w$ is defined as: 
\begin{equation} \label{eq:w}
w(\vch{e},\vc{r})
  \equiv \frac{1}{r^2} w_b(\vch{e},\vch{r}) w_\eps(\vc{r}),
\end{equation}
where $w_b$ is an instrument beam, pointed at the direction $\vch{e}$, which depends on angular coordinates $\vch{r}$, 
while $w_\eps$ is a window function defining the extent of the observed object.

The Fourier transform of a spectral line is
\begin{equation} \label{eq:sig_ft}
\begin{array}{ll}
\ft{S}(\vch{e},k_v) 
  & \equiv \frac{1}{2\pi} \int_{-\infty}^\infty S(v_0) e^{-i k_v v_0} \D{v_0} \\
  & = \ft{f}(k_v) \int w(\vch{e},\vc{r}) \D{\vc{r}} \cdot \\
  &   \eps(\vc{r}) \exp(-i k_v (v_r(\vc{r}) + v_r^{reg}(\vc{r}))).
\end{array}
\end{equation}

If we correlate $\ft{S}$ in  two directions, pointed by $\vch{e}_1$ and $\vch{e}_2$, we get the following measure
\begin{equation} \label{eq4} 
\begin{array}{ll}
K(\vch{e}_1,\vch{e}_2,k_v) 
  & \equiv \avg{\ft{S}(\vch{e}_1,k_v) \ft{S}^*(\vch{e}_2,k_v)} \\ 
  & =\ft{f}^2(k_v) \int w(\vch{e}_1,\vc{r}) \D{\vc{r}} \int w(\vch{e}_2,\vc{r'}) \D{\vc{r'}} \cdot\\
  & \avg{\eps(\vc{r})\eps(\vc{r'})} \avg{\exp(-i k_v (v_r(\vc{r})-v_{r'}(\vc{r'})))}  \cdot\\
  & \exp(-i k_v (v_r^{reg}(\vc{r})-v_{r'}^{reg}(\vc{r'}))),
\end{array}
\end{equation}
where $\langle...\rangle$ denotes averaging and where we assumed that gas velocity and emissivity are uncorrelated.

Equation \ref{eq4}  gives us the emissivity correlation function: 
\begin{equation} 
C_\eps(\vc{r}-\vc{r'})=\avg{\eps(\vc{r})\eps(\vc{r'})}.
\end{equation}

We provide averaging assuming that the velocity statistics are Gaussian:\footnote{We assume that the velocity field is Gaussian. The latter is true to high accuracy (see LP00).}
\begin{equation} 
\begin{array}{ll}
&\avg{\exp(-i k_v (v_r(\vc{r})-v_{r'}(\vc{r'})))} \\
&  = \exp\left(-\frac{k_v^2}{2} \avg{(v_r(\vc{r})-v_{r'}(\vc{r'}))^2}\right).
\end{array}
\end{equation}
We assume that the beam separation and the beam width are both small enough that we can neglect the difference between 
$v_r$ and $v_z$ (we consider $z$-axis to be a bisector of the angle between beams). 
We also assume that $v_z^{reg}(\vc{r})$ depends only on $z$:
\begin{equation} \label{eq:uz_lin} 
v_z^{reg}(z)
  = b(z-z_0)+v_{z,0}^{reg},
\end{equation}
where $b$ characterizes regular velocity shear. 
The case of velocity shear $b$ arising from Galaxy rotation is considered in detail in LP00.

Let us introduce the velocity structure function:
\begin{equation}
D_{vz}(\vc{r}-\vc{r'})
  \equiv \avg{(v_z(\vc{r})-v_z(\vc{r'}))^2}, 
\end{equation}
then Eq.~(\ref{eq4}) can rewritten as:
\begin{equation} \label{eq:K12_inter} 
\begin{array}{ll}
K(\vch{e}_1,\vch{e}_2,k_v)
  & =\ft{f}^2(k_v) \cdot \\
  & \int w(\vch{e}_1,\vc{r}) \D{\vc{r}} \int w(\vch{e}_2,\vc{r'}) \D{\vc{r'}} C_\eps(\vc{r}-\vc{r'}) \cdot \\
  & \exp \left(-\frac{k_v^2}{2} D_{vz}(\vc{r}-\vc{r'}) - ik_v b(z-z') \right). 
\end{array}
\end{equation}

Further transformations result in
\begin{equation} \label{eq:K12} 
\begin{array}{ll}
K(\vch{e}_1,\vch{e}_2,k_v) 
  & =\ft{f}^2(k_v) \int g(\vch{e}_1,\vch{e}_2,\vc{r}) \D{\vc{r}} \cdot\\
  & C_\eps(\vc{r}) \exp \left(-\frac{k_v^2}{2} D_{vz}(\vc{r}) - ik_v b z \right), 
\end{array}
\end{equation}
where geometric factor $g$ is defined
\begin{equation}
g(\vch{e}_1,\vch{e}_2,\vc{r}) \label{eq:g_orig}
  \equiv \int w(\vch{e}_1,\vc{r'})w(\vch{e}_2,\vc{r'}+\vc{r}) \D{\vc{r'}}, 
\end{equation} 
and $\ft{f}$ is a Fourier transform of the channel sensitivity function $f$.

It is easy to see that the measure $K$ given by Eq.~(\ref{eq:K12}) depends both on the angular distance between the vectors $\vch{e}_1$ and $\vch{e}_2$ 
and the velocity wavevector $k_v$. If we integrate over $k_v$ 
we are left with the 2D power spectrum that is the basis of the VCA technique. On the other hand, 
if we take coincident vectors $\vch{e}_1$ and $\vch{e}_2$ to obtain a one-dimensional spectrum $P_1$, we arrive at the VCS technique that we employ in this paper.

\begin{equation} \label{eq:P1_orig} 
\begin{array}{ll}
P_1(k_v)
  & \equiv K(\vch{e}_1,\vch{e}_2,k_v)\vert_{\vch{e}_1=\vch{e}_2} 
  = \ft{f}^2(k_v) \int g(\vc{r}) \D{\vc{r}} \cdot \\
  & C_\eps(\vc{r}) \exp \left(-\frac{k_v^2}{2} D_{vz}(\vc{r}) - i k_v b z \right).
\end{array}
\end{equation}
where $g(\vc{r}) \equiv g(\vch{e},\vch{e},\vc{r})$ and $\vch{e}$ is a beam direction.

Eq. (\ref{eq:P1_orig}) has two asymptotic spectral regimes that depend on beam size (see LP00 and CL06) and on whether the density spectral slope is shallower than -3 or steeper than -3.  The gist of our approach  is that we
do not use the asymptotic regimes but rather calculate P1 directly.

First of all, we derive the window function. SMC is a remote emitting structure and therefore the corresponding window function $w_\eps$ is non-zero only in some vicinity of a distant enough point $z_0$. We also assume, that picture-plane extent of the object is larger than the beam size, so the $\vc{R}$-dependence of $w_\eps$ can be ignored. Then we obtain
\begin{equation}
w(\vc{r})
  \approx \frac{1}{z_0^2} w_b(\theta) w_\eps(z)
\end{equation}
If we introduce
\begin{equation} \label{eq:Wb_plos_def}
W_b(R)
  \equiv \frac{1}{z_0^2} w_b\left(\frac{R}{z_0}\right),
  \qquad \smallint W_b(R) \D{\vc{R}} = 1
\end{equation}
\noindent window function becomes
\begin{equation}
w(\vc{r})
  = W_b(R) w_\eps(z)
\end{equation}
and the geometric factor $g(\vc{r})$ is the auto-correlation of it. In our calculations we use Gaussian approximation for both $W_b(R)$ and $w_\eps(z)$. 

Our goal is to express $D_{vz}$ through the velocity spectrum. If we assume that $\vc{v}(\vc{r})$  is primarily solenoidal with power-law power spectrum having cutoff at large scales, the velocity power spectrum can be written in as follows (see, for example, \citealt{Les91}):
\begin{equation} \label{eq:Fu} 
F_{ij}(\vc{k})
  = \frac{V_0^2}{k^{\alpha_v}} e^{-\frac{k_0^2}{k^2}} \left(\delta_{ij} - \frac{k_i k_j}{k^2}\right),
\end{equation}
\noindent where $V_0^2$ is the velocity power spectrum amplitude, $\alpha_v$ is the velocity spectral index, $k_0=2\pi/L_v$ is the cutoff wavevector bound with the injection scale $L_v$, and  $i$ and $j$ are the component indexes. Then $D_{vz}$ can be represented as
\begin{equation} \label{eq:Dz_Fu} 
D_{vz}(\vc{r})
  = 2 \int \D{\vc{k}} (1 - e^{i\vc{k}\vc{r}}) \hat{z}_i\hat{z}_j F_{ij}(\vc{k}),
\end{equation} 
where the summation over repeating indexes is assumed.  We point out that Eq. \ref{eq:Fu}  contains the injection scale $L_v=2\pi/k_0$
and the velocity power spectrum amplitude $V_0^2$. Thus the injection energy and injection scale can also be found from the VCS in addition the power spectral index.

%\subsection{Emissivity correlation function}

In addition, we need to estimate the emissivity correlation function for our modeling. 
This part can be presented as a sum of variable and constant parts:
\begin{equation}
C_\eps(\vc{r}) = \int A_\eps^2 F_\eps^2(\vc{k}) e^{i\vc{k}\vc{r}} \D{\vc{k}} + \eps_0^2  
\end{equation}
where $A_\eps^2 F_\eps^2(\vc{k})$ is the emissivity power spectrum and $\eps_0$ is the mean emissivity. Here $A_\eps^2$ is the spectrum amplitude.

For our modeling we need the value of $\delta_\eps^2 = A_\eps^2/\eps_0^2$.
After some algebra we can derive for it the following expression:
\begin{equation}
\delta_\eps^2 = \left(\frac{\Delta S}{S_0}\right)^2  
   \left(\int |\ft{W}_{b,0}(\vc{K})|^2 |\ft{w}_{\eps,0}(k_z)|^2 F_\eps^2(\vc{k})\D{\vc{k}}\right)^{-1} 
\end{equation}
where $\Delta S/S_0$ is the relative deviation plane of sky intensity, that is equal to $0.402$ for the SMC intensity map within the selected region. $\ft{W}_{b,0}$ and $\ft{w}_{\eps,0}$ are the window function Fourier transforms normalized to their values at zero wavevector. Then we can evaluate the emissivity correlation function as follows:
\begin{equation}
C_\eps(\vc{r}) \sim \int \delta_\eps^2 F_\eps^2(\vc{k}) e^{i\vc{k}\vc{r}} \D{\vc{k}} + 1  
\end{equation}
where
\begin{equation}
F_\eps^2(\vc{k}) = \frac{e^{-\frac{k_0^2}{k^2}}}{k^{\alpha_\eps}}  
\end{equation}
For our calculations we assume that $k_0$ is the same as in Eq. \ref{eq:Fu}. 
In what follows, we use the mathematical formalism above to analyze the observational data.

\section{Results}
\label{sec:res}

The two-dimensional spatial power spectrum of the HI column density image was performed by Stanimirovic \& Lazarian (2001).  They found that
$\alpha_\eps \propto 3.3$. Using a distance to the SMC of 60kpc and thickness (size) of the SMC 4kpc (SX99) we apply the VCS procedure as outlined in the previous section. We first calculate the power spectrum over the velocity coordinate $P_1$ for three resolutions: $0^\circ.025$,  $0^\circ .05$ and $0^\circ .1$. The resulting spectra are shown in Figure \ref{fig:vcs}.  We note that our analysis is mostly biased towards the cold gas in the SMC and that the spectrum measured here is generally of the cold gas, which tends to have higher sonic Mach numbers than the warm HI in the SMC (Burkhart et al. 2010). The fitting of the spectrum, which is overplotted in Figure \ref{fig:vcs},  is performed with Mathematica \verb|FindMinimum|, which implements the algorithm of steepest descent.   

Applying the VCS to the SMC data we report the following parameters:
\begin{itemize}
\item  $\alpha_v = -3.85 \pm 0.02$
\item $\alpha_\eps = -3.3$ (Stanimirovic \& Lazarian 2001)
\item $L_v = 2.3^{+0.7}_{-0.4}$ kpc 
\item $v_{turb} = 6.7^{+1.8}_{-0.4}$ km/s
\item $M_{cold} = 5.6^{+0.8}_{-0.2}$
\end{itemize}

The mach number for the cold phase $M_{cold}$ is calculated assuming the cold phase temperature $T_{cold} = 173$K.
We note that the $P_1$ fitting did not include the first 3 points, as they are affected by the regular motions (see below).

\begin{figure*}[tbh]
\centering
\includegraphics[scale=1.2]{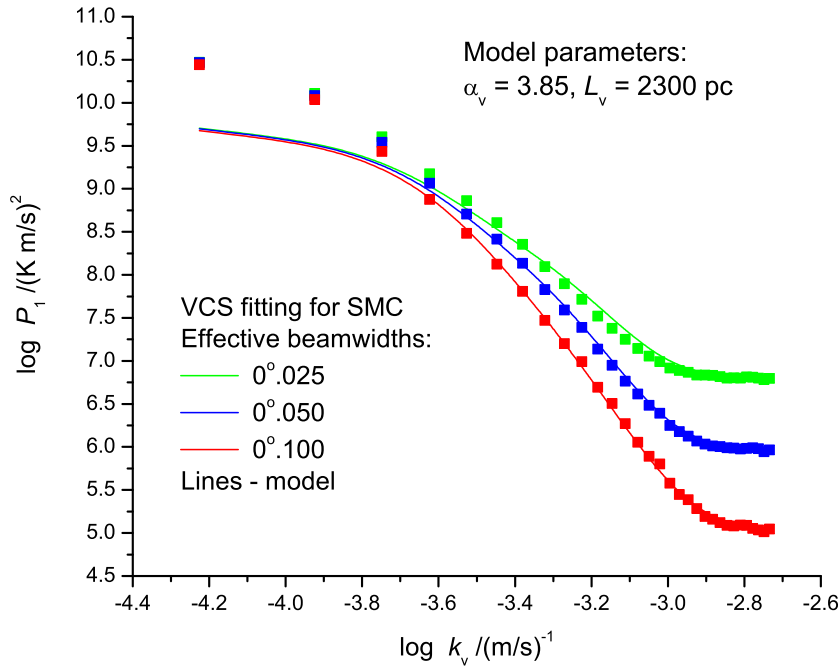}
\caption{Fitting of the model VCS spectra to $P_1$ of the SMC for different effective resolutions. We do not include in our fitting the first three points as they are affected by the regular motions.}
\label{fig:vcs}
\end{figure*}

\subsection{SMC Regular Motions and the VCS }

An important consideration for the analysis of velocity statistics in a galaxy such as the SMC is the contribution to the turbulence power spectrum from regular motions. Regular motions in the SMC were studied in Stanimirovic, Staveley-Smith \& Jones (2004) who found a complicated velocity field resulting from several expanding HI shells in addition to the galactic rotational velocity.

Let us estimate the possible impact of these motions. The velocity term of the VCS for arbitrary regular velocity is:
\begin{equation}
\exp(-\frac{k_v^2}{2}D_z(\vc{r}-\vc{r}')-ik_v(v_z^{reg}(\vc{r})-v_z^{reg}(\vc{r}')))
\end{equation}
Comparing the arguments for the turbulent and regular velocities and accounting for the fact that the spatial dependency of $D_z$ is three-dimensional,  while that of $v_z^{reg}$ is one-dimensional (due to the relatively narrow beam) we can estimate the boundary $k_v^{bound}$ which separates regular motion domination region from the region where turbulent motions dominate as follows:
\begin{equation}
k_v^{bound}=\left(\frac{\Delta v_z^{reg} }{v_{turb}^6}\right)^{\frac{1}{5}}
\end{equation}
We estimate $\Delta v_z^{reg}$ as the separation between large scale regular components of the average line-of-sight velocity.
According to our calculations, $\Delta v_z^{reg}=48000$ m/s  and $v_{turb}=6700$ m/s. 
Substituting this to the previous expression we finally have:
\begin{equation}
\log k_v^{bound}\approx -3.6
\end{equation}
We note that this estimation fits well to the approximate scale at which our VCS fitting deviates from the data.

\section{Discussion and Conclusions}
\label{sec:disc}

The velocity power spectrum  is a critical parameter of turbulence to obtain from ISM observations as it gives true dynamical insight into the process of turbulence.  This is not the first time the velocity power spectrum has been estimated for the SMC galaxy as Stanimirovic \& Lazarian (2001) used VCA to recover the velocity slope. While our estimate for the density spectrum in this paper agrees well with the value obtained using the VCA technique in Stanimirovic \& Lazarian (2001) our estimate of the velocity spectrum is substantially
different. Indeed, Stanimirovic \& Lazarian (2001) found the three-dimensional velocity power spectral slope to be -3.4, which, in fact, is an unusual value for MHD turbulence (see Brandenburg \& Lazarian 2013 for a review).  In contrast, we find a much steeper slope for the velocity power spectrum of -3.85.  This slope is predicted from numerical simulations of high Mach number turbulence  (e.g. Kritsuk 2007; Kowal \& Lazarian 2010; Collins et al. 2012) as well as the sonic Mach number analysis of the SMC  done in  Burkhart et al. (2010).
The steep velocity slope in a compressible medium also harkens to the predictions of a steep velocity slope given in the Burgers shock dominated turbulence model.

We believe that fitting of integral expressions that we employed here provides a more accurate approach compared to the asymptotic approach applied in Stanimirovic \& Lazarian (2001). The observed spectrum of fluctuations is limited by the resolution of the data and, given the limits of sensitivity, the uncertainty of the spectral index evaluation may be substantial. For instance, if we consider measuring the slope of the spectra of 2D channel maps within VCA, we should be aware for a given channel thickness the slices may show ``thick'' slice asymptotics for smallest eddies, while for large eddies the slices will exhibit ``thin'' asymptotics. Thus the spectrum at small 
wavenumber may have a spectral slope which  differs from that of large wavenumbers and fitting a single power law may be misleading. A possible approach to dealing with this issue is choosing the channel thickness of the order of the thermal line width. This however may require much higher signal-to-noise  data. As a result, a similar procedure of fitting the spectra by using integral expressions and varying the thickness of
the channel maps may be useful for the VCA practical analysis of the data. We believe that this should reconcile the data of the VCA with our results. Nevertheless, we want to point out that the compressible and incompressible components of velocity field contribute differently into the measures available to the VCA and VCS. This may be another reason for the disagreement that we face. The issue requires further studies.    

What drives the turbulence in the SMC? The fitted value of the 2.3 kpc injection scale suggests two dominate possibilities. 
One possibility is that turbulence is stirred 
 by a large number of expanding HI shells found in the
SMC. The shell sizes range from $\approx$30 pc to $\approx$2 kpc. Hence,
one scenario could be that the largest shells drive the turbulent
cascade down to the smallest observed scales or that the shells fragment 
and drive instabilities.  Turbulence may also be driven in the SMC by the last SMC-LMC
encounter.  This would agree with the predictions made by simulations
presented in Besla et al. (2012) and also with the analysis of Goldman (2000).
We did not find evidence for a dissipation scale present in the HI data however the scales resolved in our data only proceed down to 30pc and thus the dissipation scale in the SMC may be smaller than this scale.  This agrees with the line width size relation of the SMC, which as been observed down to scales smaller than 30pc (Rubio, Lequeux \& Boulanger 1993).

In light of the VCS analysis we can estimate the energy density of turbulence in the SMC.
Given a turbulent velocity at the injection scale of v$_{turb}$=6700 m/s  and an average ambient  HI density of 0.3-1.9 atoms/cm$^{3}$ (SX99), then the energy density of turbulence in the SMC would range between
$w_{turn}=0.07-0.44$ eV/cm${^3}$, as compared to $\approx 1$ev/cm${^3}$ for the Milky Way.
As a comparison, the energy density in cosmic ray electrons in the SMC has been estimated by the Fermi telescope to be
$w_{CR_e}$=0.0036  eV/cm${^3}$, which is about 60\% of the local cosmic ray energy density in the  Milky Way (Abdo et al. 2010),  and the energy density in the cosmic ray protons is estimated to be
$w_{CR_p}$=0.15  eV/cm${^3}$ (Abdo et al. 2010).
Mao et al. 2008 estimated a total 3D field strength in the SMC (assuming equipartition with cosmic rays) of 1.7$\pm0.4 \mu G$ which 
suggests a magnetic energy density of $\approx$0.12$\pm$0.06 ev/cm${^3}$.
%IThis suggests a plasma beta in the diffuse SMC of 
%I$\beta=P_{gas}/P_{mag}$=8$\pi nkT/B^2\approx 5.2$, given T=3800K and n=1.9cm${^-3}$.
%IFor the dense SMC HI supershells, this implies that the cosmic rays may not be confined efficiently.
%IUsing the lower range of density or lower values for temperature in the SMC may suggest a $\beta<1$ thus we conclude 
%Ithe diffuse gas in the SMC has plasma $\beta \approx 1$ given the uncertainty in the measured parameters and the fact that a range of reasonable parameters does not give us a value much larger than unity.
The Milky Way appears to have an equipartition of energy between turbulence, magnetic fields, and cosmic rays with each component carrying about 1 ev/cm$^{3}$.  The SMC is a dwarf galaxy (with dynamical mass of 2.4$\times 10^9$ M$_{\odot}$) and thus it is unsurprising that the overall energy density in turbulence, magnetic fields, and cosmic rays, is significantly less. 

The VCS technique, which as been numerically tested in Chepurnov et al. 2009,  has a bright future ahead in terms of measuring the properties of turbulence
in a number of interesting environments.  The technique thus far has been applied primarily in the context of 21 cm data but there is no reason not to apply it to the wealth of molecular line tracers such as CO as was done in Padoan et al. (2009).  Accurate measurements of the turbulence energy budget of molecular clouds and the sonic Mach number will have important  consequences for theories of star formation (i.e. see Krumholz \& McKee 2005; Padoan \& Nordlund 2002; Hennebelle \& Chabrier 2008; Federrath \& Klessen 2012, and references therein).
The VCS does not need to be limited to ISM or galaxy studies but also can be applied to spectral lines as measured in galactic halo environments or galaxy clusters using absorption lines or x-ray emission lines with high enough spectral resolution.

Our results in this paper are in agreement with the study of Mach numbers of turbulence in SMC in Burkhart et al. (2010). This illustrates well the synergy of using different techniques to investigate the parameters of  turbulence. A new technique of studying turbulence spectra of magnetic fluctuations using synchrotron polarization fluctuations was recently developed in Lazarian \& Pogosyan (2015). This technique makes use  of a statistical description of synchrotron statistics and applicable to describing fluctuations for arbitrary spectral index of cosmic rays. It will be advantageous to study and compare the statistics of velocity and magnetic field in order to obtain a comprehensive picture of turbulence in SMC and other galaxies. 

% This may again point to evidence that the LMC and SMC have already interacted and cosmic rays have diffused rapidly away from the SMC.
%The LMC cosmic ray proton energy density was  estimated  by Foreman et al. (2015) and  Abdo et al. (2010c) 
%to be $w_{CR_p}$=0.199-0.31 eV/cm^{-3}. 

\section{Conclusions}
\label{sec:con}

We apply the Velocity Coordinate Spectrum technique to obtain the velocity power spectrum in the SMC galaxies in 21cm emission.  We find a steep power spectral slope for velocity which is in agreement with the expectations for supersonic turbulence.
In particular, we obtain the following properties of the turbulence velocity spectrum in the diffuse gas of the SMC:
\begin{itemize}
\item The velocity spectral slope: $\alpha_v = -3.85 \pm 0.02$
\item The density spectral slope: $\alpha_\eps = -3.3$ (Stanimirovic \& Lazarian 2001)
\item The injection scale of turbulence: $L_v = 2.3^{+0.7}_{-0.4}$ kpc 
\item The turbulence velocity amplitude: $v_{turb} = 6.7^{+1.8}_{-0.4}$ km/s 
\item Mach number for the cold phase $M_{cold} = 5.6^{+0.8}_{-0.2}$
\end{itemize}

\acknowledgments
B.B. is grateful for support from the NASA Einstein Fellowship.
A.L. and B.B. thank the Center for Magnetic Self-Organization in Astrophysical and Laboratory Plasmas for financial support. AL is supported by the NSF grant AST 1212096.
S.S. acknowledges support by the NSF Early Career Development (CAREER) Award AST-1056780.

\end{document}